\documentclass[3p,times,twocolumn]{elsarticle}

\usepackage{ecrc}

\volume{00}

\firstpage{1}

\journalname{Nuclear Physics B Proceedings Supplement}

\runauth{}

\jid{nuphbp}

\jnltitlelogo{Nuclear Physics B Proceedings Supplement}  

\usepackage{amssymb,amsmath,float}

\usepackage[figuresright]{rotating}

\begin{document}

\begin{frontmatter}



\dochead{}  

\title{Enrico Fermi and the Dolomites}


\author{Giovanni Battimelli}
 \ead{giovanni.battimelli@uniroma1.it}
\address{Universit\`a di Roma ``La Sapienza'', Roma, Italy}

\author{Alessandro de Angelis}
 \ead{alessandro.de.angelis@cern.ch}
\address{Universit\`a di Udine and INFN, Udine, Italy}

\begin{abstract}
Summer vacations in the Dolomites were a tradition among the professors of the Faculty of Mathematical and Physical Sciences at the University of Roma since the end of the XIX century. Beyond the academic walls, people like Tullio Levi-Civita, Federigo Enriques and Ugo Amaldi sr., together with their families, were meeting  friends and colleagues in Cortina, San Vito, Dobbiaco, Vigo di Fassa and Selva, enjoying trekking together with scientific discussions. The tradition was transmitted to the next generations, in particular in the first half of the XX century, and the group of via Panisperna was directly connected: Edoardo Amaldi, the son of the mathematician Ugo sr., rented at least during two summers, in 1925 and in 1949, 
and in the winter of 1960, a house in San Vito di Cadore, and almost every year in the Dolomites;   Enrico Fermi was a frequent guest. Many important steps in modern physics, in particular the development of the Fermi-Dirac statistics and the Fermi theory of beta decay, are related to scientific discussions held in the region of the Dolomites.
\end{abstract}


\end{frontmatter}



Between the end of the XIX century and the '50s, the Dolomite mountains have been a traditional vacation destination for the mathematicians and the physicists of the University of Roma, together with their families. The tradition begun with a small community of mathematicians, among which Tullio Levi Civita (1873 - 1941), Federigo Enriques (1871 - 1946), Guido Castelnuovo (1865 - 1952) and Ugo Amaldi sr. (1875 - 1957). Among the destinations were Cortina d'Ampezzo, San Vito di Cadore, Dobbiaco, Vigo di Fassa and Selva di Val Gardena, where the professors and their families could enjoy trekking together with scientific discussions \cite{battimelli2003,battimelli2004}.

The tradition was transmitted to the next generations and the younger scientists added to the predilection of the fathers for the long stays in the mountains a passion for physical exercise, which led them to a more active practice of mountain sports.

The group of via Panisperna was directly connected to this tradition: Edoardo Amaldi (1908 - 1989), the son of the mathematician Ugo sr.,  and  General Secretary of CERN\footnote{This was in the first years of CERN the title of the charge now called Director General.} in its provisional phase (1951 - 1954), rented at least during two summers, in 1925 and in 1949, and in the winter of 1960, a house in San Vito di Cadore, and almost every year in the Dolomites;   Enrico Fermi 
(1901 - 1954) was a frequent guest.

\begin{figure}[h]
\center{\includegraphics[width=\columnwidth]{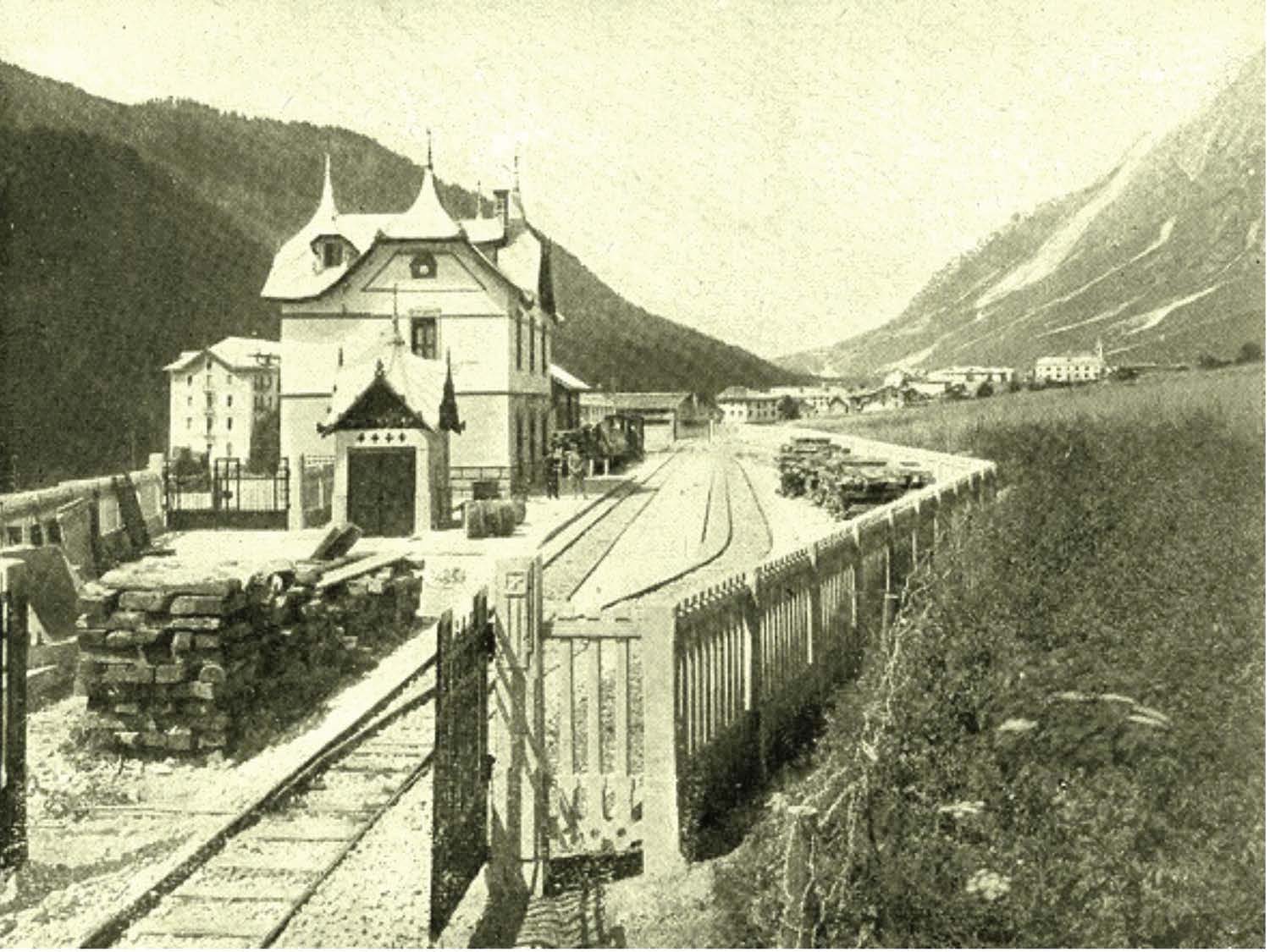}}
\caption{\label {cartolina} San Vito di Cadore in the '20s of last century.}
\end{figure}

In the '20s of last century Enrico Fermi  was in  San Vito di Cadore, a guest of Taresina Menegus, from Ruseco; later he rented a house in Cortina, and in 1925 he came  to San Vito to meet the Amaldi family who were renting  part of the house of Tita Menegus, another member of Menegus family.Ê One could hear ``tutte le sante notti'' (every holy night) the noise of hand-cranked mechanical calculators \cite{Pordon2013}.

In 1925 the young Edoardo Amaldi  spent several days with Enrico Fermi, then interim professor of theoretical mechanics in Firenze, during a long bike ride on the Dolomites. A deep friendship was born, and the interest of Edoardo Amaldi for physics.

\begin{figure}[h]
\center{\includegraphics[width=\columnwidth]{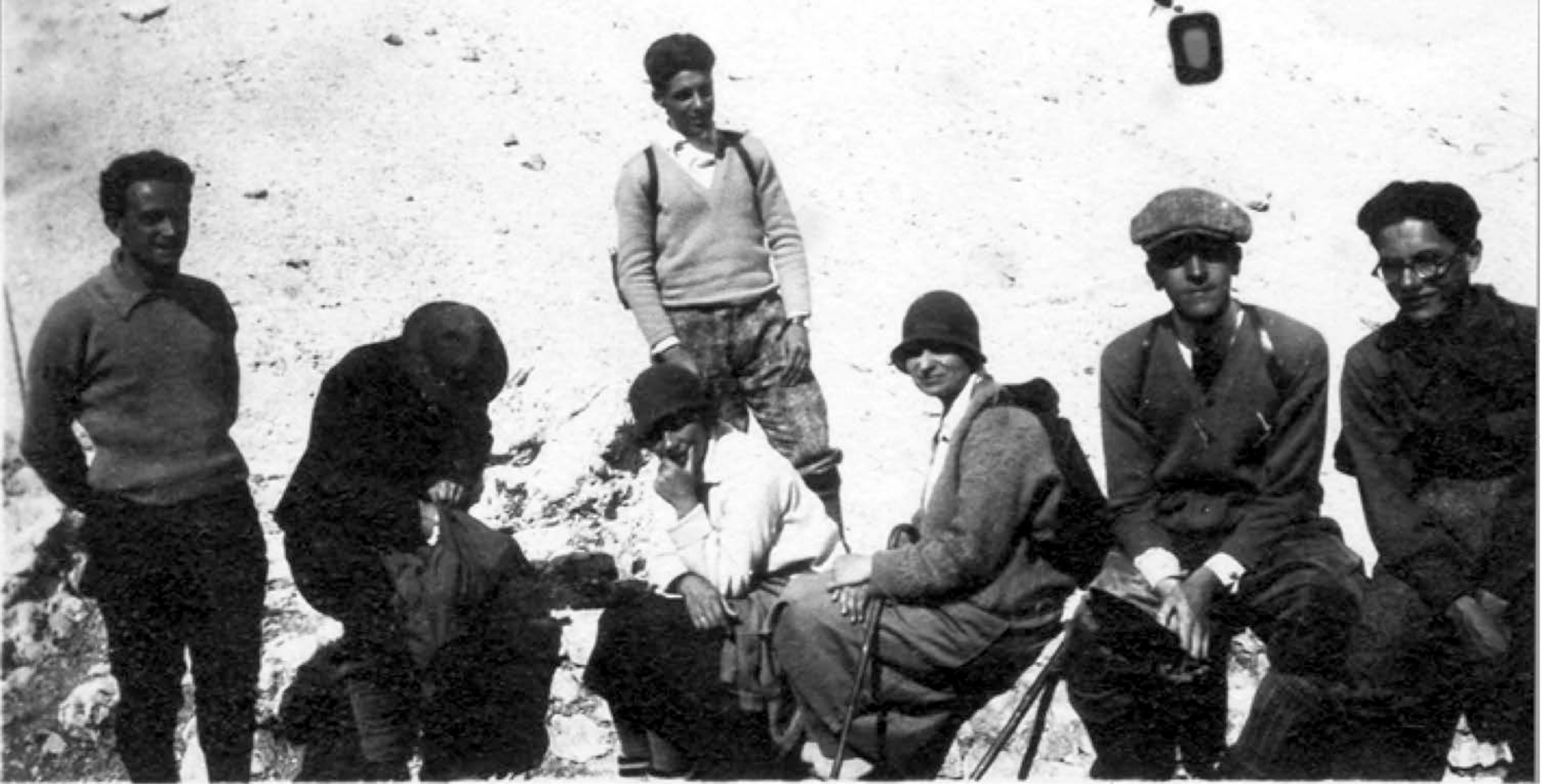}}
\caption{\label {1926} Summer 1926: Fermi (left) and E. Amaldi (right) during an excursion.}
\end{figure}

The summer of 1925 in the Dolomites was not only the key of a new friendship, but was also at the root of an article changing the history of physics.

In 1925 Wolfgang Pauli (1900 - 1958) announced his exclusion principle. Fermi invited in the Dolomites his friend Ralph Kronig (1904 - 1995), a young brilliant physicist, who had just concluded
his PhD thesis and joined the company. In January 1925 Kronig had first proposed electron spin after hearing a seminar by Pauli in T\"ubingen (Heisenberg and Pauli first rejected the idea!). Discussing with Kronig, Fermi sketched a paper in which he applied the Pauli principle to an ideal gas, employing a statistical formulation now known as Fermi-Dirac statistics  \cite{Fermi1926}.

\begin{figure}[h]
\center{\includegraphics[width=\columnwidth]{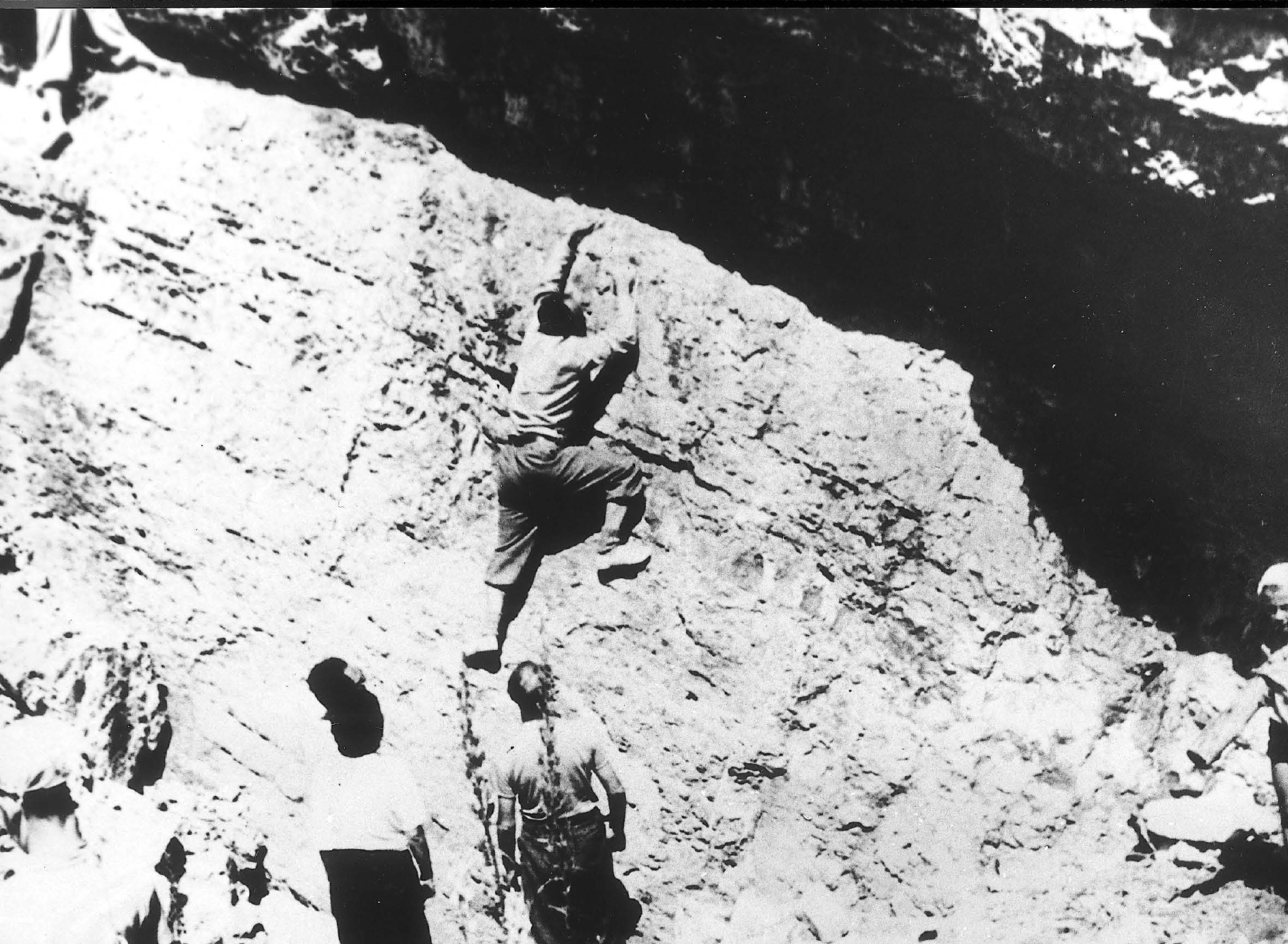}}
\caption{\label {fermiclimbs} Enrico Fermi climbing in the Dolomites.}
\end{figure}

\begin{figure}[h]
\center{\includegraphics[width=\columnwidth]{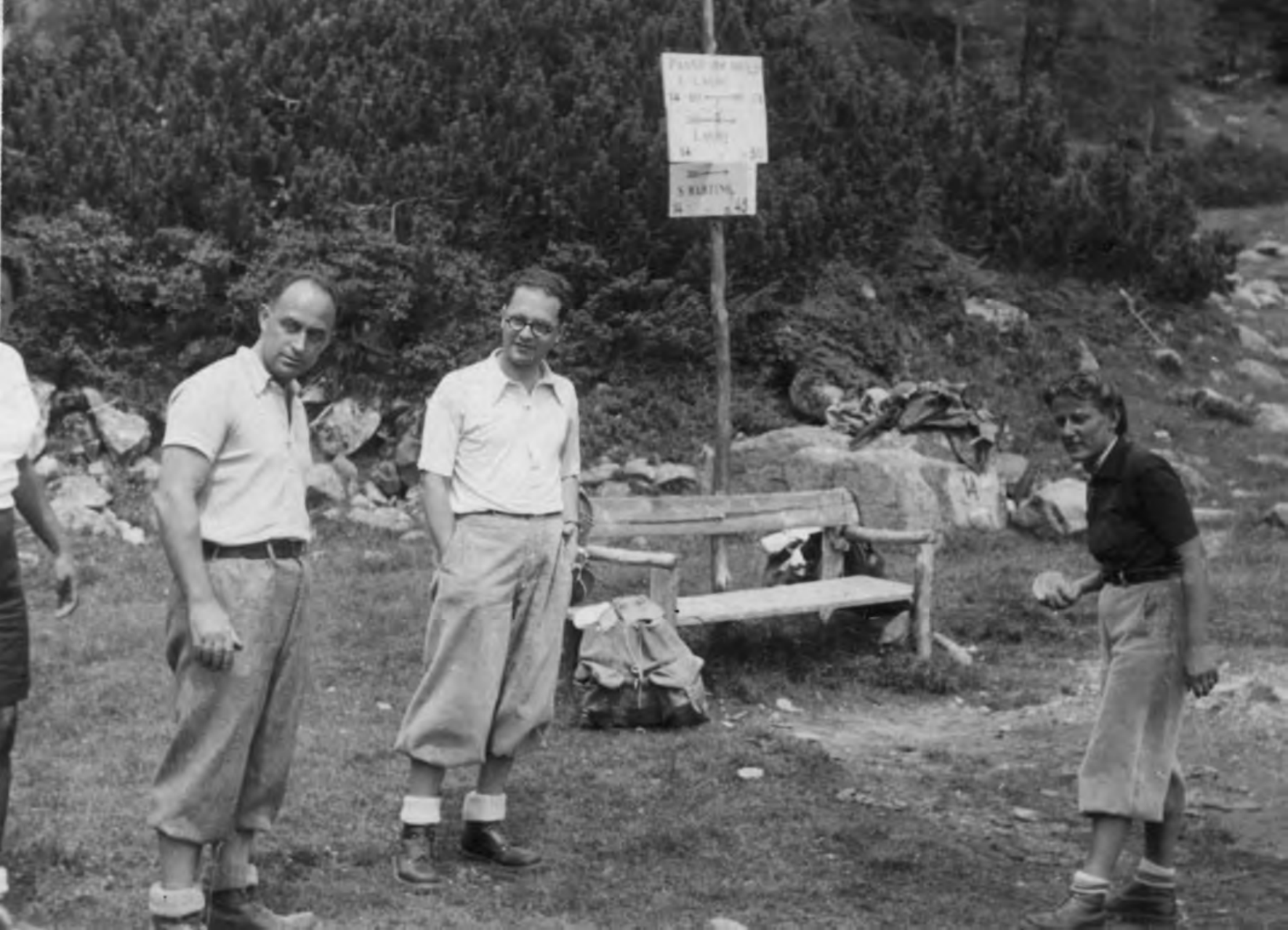}}
\caption{\label {bocce}Fermi and Edoardo Amaldi playing {\em bocce} (the Italian lawn bowling game) in San Martino di Castrozza, 1938.} 
\end{figure}

A visit to Cortina and in Val Gardena by Fermi is reported also in 1926, the year in which he was assigned the chair of theoretical physics in Roma.

Vacations in the Dolomites and scientific activity continued to blend together. By the end of 1933 Fermi conceived his theory of beta decay, which was eventually published early in 1934 \cite{5}. Emilio Segr\`e (1905 - 1989) vividly recollects how Fermi explained the basic ideas of the theory to the younger members of the team, squeezed in a tiny hotel room in val Gardena, during the Christmas holidays they were spending skiing in the Dolomites \cite{6}.

\begin{figure}[h]
\center{\includegraphics[width=\columnwidth]{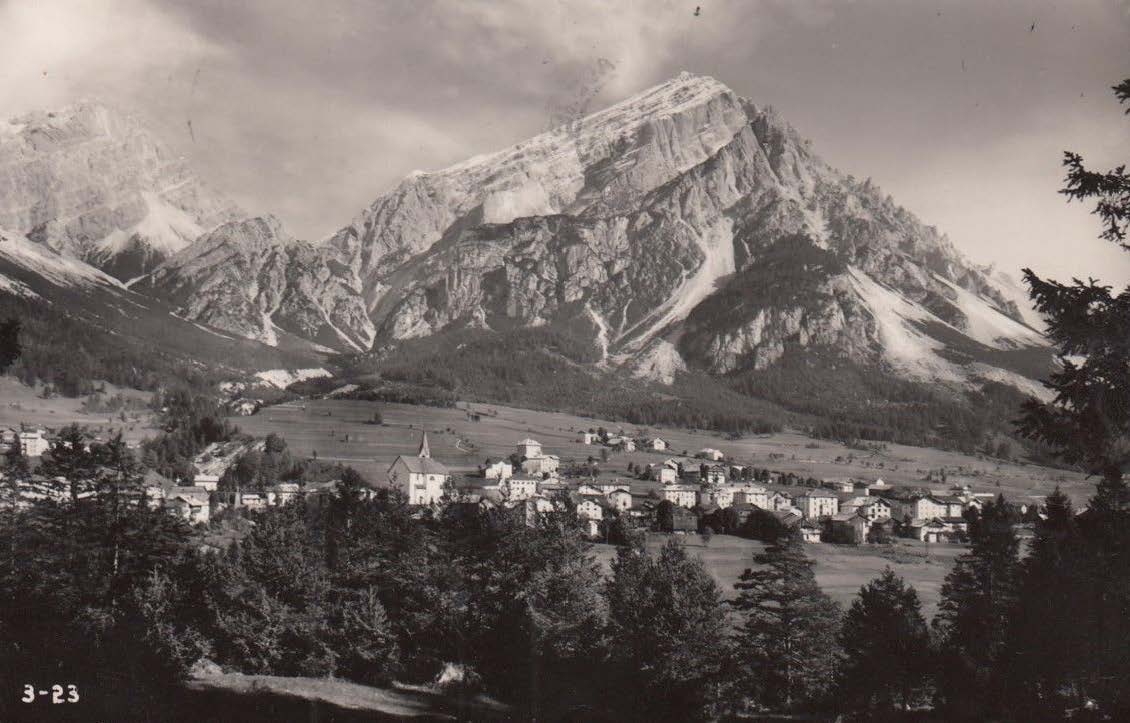}}
\caption{\label {cartolina2}San Vito di Cadore around  1950.} 
\end{figure}

\begin{figure}[h]
\center{\includegraphics[width=0.5\textwidth]{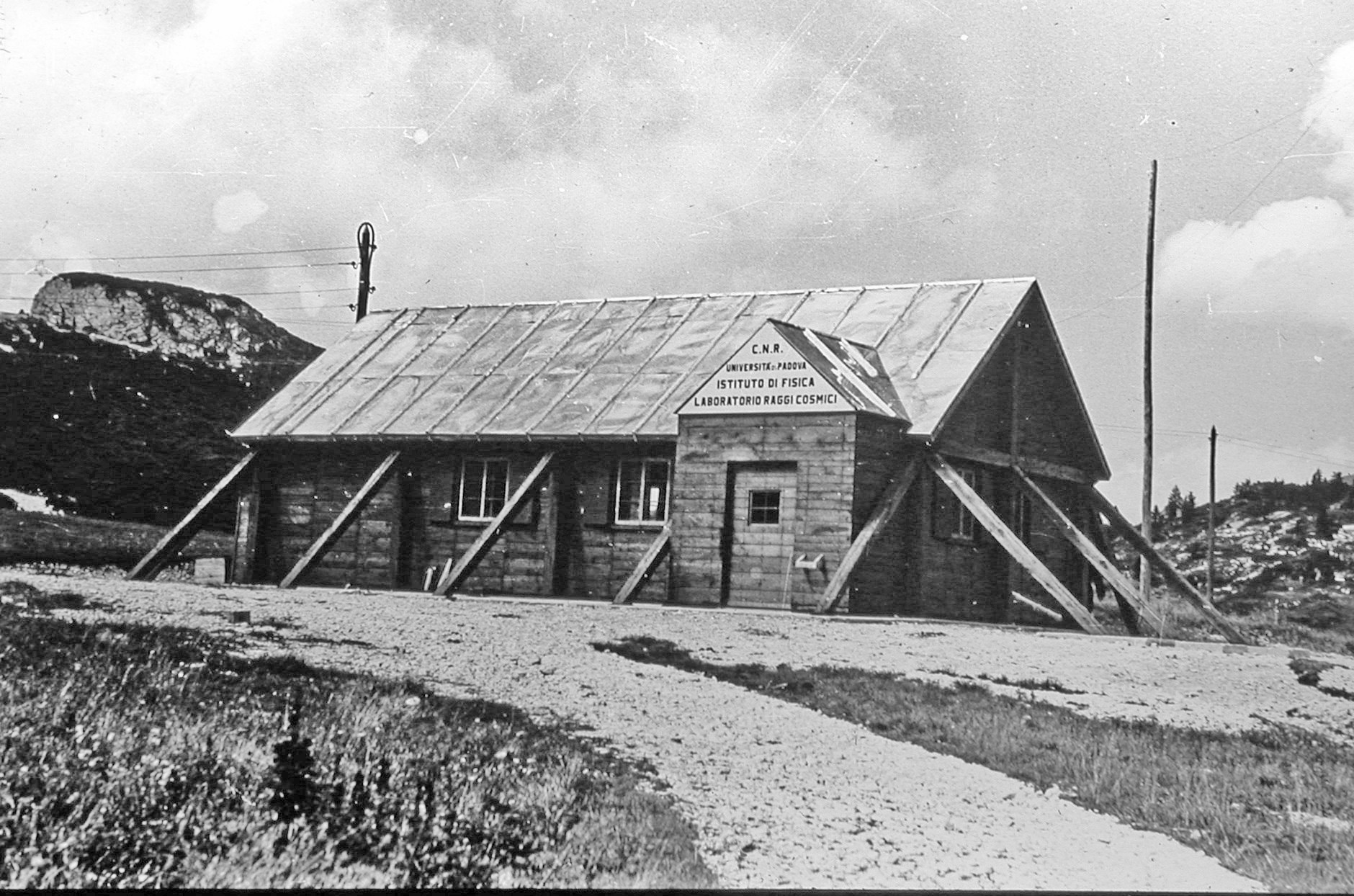}}
\caption{\label{fedaia}The Cosmic Ray laboratory in Passo Fedaia, under the Marmolada.} 
\end{figure}

\begin{figure}[h]
\center{\includegraphics[width=0.9\columnwidth]{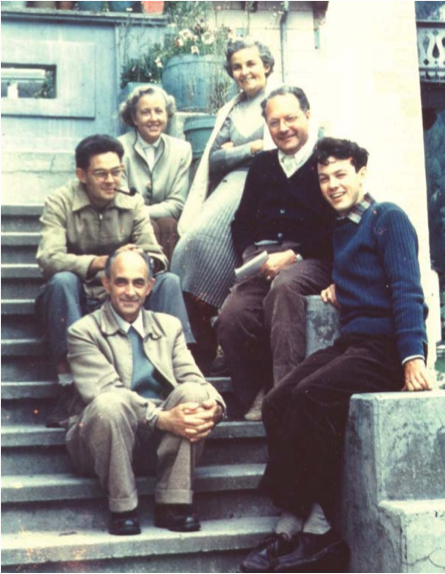}}
\caption{\label {ultimafoto}The Fermi and  Amaldi families in Pera di Fassa in 1954. The picture was taken by Enrico Persico (1900 - 1969).} 
\end{figure}

Fermi loved physical exercise and hiking in the mountains, but was not really an experienced climber, as Franco Rasetti (1901 - 2001) remembered: ``Fermi was not a mountain climber. He was extremely sturdy, very strong. He had a lot of endurance, but he was afraid of steep slopes. Anything steep scared him. He could walk thirty or forty kilometers in a day, or cycle long distances. He liked the mountains but he wouldn't climb them'' \cite{7}.

In 1938 for the last time  Fermi spent his summer vacations together with his friends and colleagues in the Dolomites before World War II. Early in September, while enjoying their holidays in San Martino di Castrozza, news reached them of the first anti-semitic laws just promulgated by the fascist government, an outcome anticipated by the publication in July of the infamous ``Manifesto della razza''.

Fermi's wife, Laura, was of Jewish family. The racial laws were the last sign of deterioration of the situation in Italy that convinced Fermi to take the decision to leave the country. At the end  of 1938 he was awarded the Nobel prize; with his family, he left Italy for Stockholm, and moved then straight to the United States. He would not come back to Italy (and to the Dolomites) for over ten years.

The famous Fermi second order acceleration mechanism (1949), providing for the first time an explanation for the production of cosmic rays, was thus conceived in the US;
however, Fermi came back 
 to Italy after 1938 for the first time in 1949, and he attended the International Conference on Cosmic Rays  in Como, where he presented  to the cosmic ray community his idea as
``Una teoria sull'accelerazione dei raggi cosmici'' (``A theory on the acceleration of cosmic rays'') \cite{Fermi1949}. On that occasion, though he did not go to the Eastern Alps, he paid a visit to the Testa Grigia, the site of one of the early cosmic-ray laboratories, built in 1947 by Italian physicists at Plateau Rosa, high above Cervinia at an altitude of 3500 meters above sea level.

Later, Fermi visited once more the Dolomites in 1954, just before his death.
After attending the summer schools in Les Houches (where he did not miss the opportunity to be carried by t\'el\'eph\'erique to the Laboratoire des Cosmiques at the Col du Midi \cite{Glauber2002}) and in Varenna, and giving as usual brilliant lectures, he spent some time together with the Amaldi family in Vigo di Fassa. On this occasion, Fermi is reported to have visited the  Cosmic Ray Laboratory in the {\em passo Fedaia}, built   in the beginning of the '50s by he University of Padova
under the dam that collects water from the glacier of Marmolada to power an electric plant \cite{DeAngelis2012a,DeAngelis2012}. Thanks to the availability of electric power, an electromagnet especially designed by the engineer Giovanni Someda 
(1901-1978) could be used to separate the particles with different electrical charges; the laboratory was partly equipped with instruments built by Bruno Rossi (1905 - 1993). Visits by 
Patrick Blackett (1897 - 1974) and Cecil Powell (1903 - 1969) to this laboratory are also reported. The building still exists.



\vspace*{2mm}
{\textit{Acknowledgements:}} The authors thank Ugo Amaldi, Adele La Rana and Barbara De Lotto for reviewing the document. AdA thanks Francesco Pordon for oral information, and for his kind hospitality in San Vito di Cadore.









\end{document}